\documentclass[aps,amsfonts,amsmath,prd,preprint,nofootinbib]{revtex4}
\usepackage{graphicx}

\newcommand{\beq}{\begin{equation}}
\newcommand{\eeq}{\end{equation}}

\def\lap{\lower.5ex\hbox{$\; \buildrel < \over \sim \;$}}
\def\gap{\lower.5ex\hbox{$\; \buildrel > \over \sim \;$}}

\begin{document}

\title{Non-singular bounce transitions in the multiverse}

\author{Jaume Garriga$^a$, Alexander Vilenkin$^b$ and Jun Zhang$^b$}

\address{$^a$ Departament de Fisica Fonamental i Institut de Ciencies del Cosmos,\\
Universitat de Barcelona, 
Marti i Franques, 1, 08028, Barcelona, Spain\\
$^b$Institute of Cosmology, Department of Physics and Astronomy,\\ 
Tufts University, Medford, MA 02155, USA}

\begin{abstract}

{According to classical GR, negative-energy (AdS) bubbles in the multiverse terminate in big crunch singularities.  
It has been conjectured, however, that the fundamental theory may resolve these singularities and replace them by non-singular bounces.  Here we explore possible dynamics of such bounces using a simple modification of the Friedmann equation, which ensures that the scale factor bounces when the matter density reaches some critical value $\rho_c$.  This is combined with a simple scalar field `landscape', where the energy barriers between different vacua are small compared to $\rho_c$.  We find that the bounce typically results in a transition to another vacuum, with a scalar field displacement $\Delta\phi \sim 1$ in Planck units.
If the new vacuum is AdS, we have another bounce, and so on, until the field finally transits to a positive-energy (de Sitter) vacuum.
We also consider perturbations about the homogeneous solution and discuss some of their amplification mechanisms (e.g., tachyonic instability and parametric resonance).  For a generic potential, these mechanisms are much less efficient than in models of slow-roll inflation.  But the amplification may still be strong enough to cause the bubble to fragment into a mosaic of different vacua.}

\end{abstract}

\maketitle

\section{Introduction}

Inflationary cosmology has profound implications for the global structure of the universe.  
If the underlying particle theory admits a number of metastable vacuum states, it leads to the picture of a multiverse, where different spacetime regions are occupied by different vacua.  Transitions between the vacua can occur through quantum tunneling, with bubbles of daughter vacuum nucleating and expanding in the parent vacuum background.\footnote{Transitions between vacua can also occur through quantum diffusion.  Here we assume for simplicity that bubble nucleation is the only mechanism.}  As a result, the entire landscape of vacua is explored. 

The spacetime structure of an eternally inflating multiverse, as it is usually assumed, is schematically illustrated in a causal diagram in Fig.1. The bubbles expand, rapidly approaching the speed of light, so the worldsheets of the bubble walls are well approximated by light cones. Disregarding quantum fluctuations, bubble interiors are open FRW universes. If the vacuum inside a bubble has positive energy density, the evolution is asymptotically de Sitter, and the bubble becomes a site of further bubble nucleation. The horizontal line at the top of the diagram represents the spacelike future infinity of such de Sitter (dS) bubbles.

Negative-energy (anti-de Sitter, or AdS) vacua, on the other hand, collapse to a big crunch and develop curvature singularities, which are represented by zig-zag lines in the figure.  It is assumed that spacetime terminates at these singularities.\footnote{Some supersymmetric particle physics models, including string theory, also predict the existence of stable Minkovski vacua.  Such vacua are not relevant for our discussion in this paper, so we do not consider them here.}  AdS bubbles are referred to as Òterminal bubblesÓ, because inflation comes to an end there and no further bubble formation takes place.  

\begin{figure}[t]
\begin{center}
\includegraphics[width=12cm]{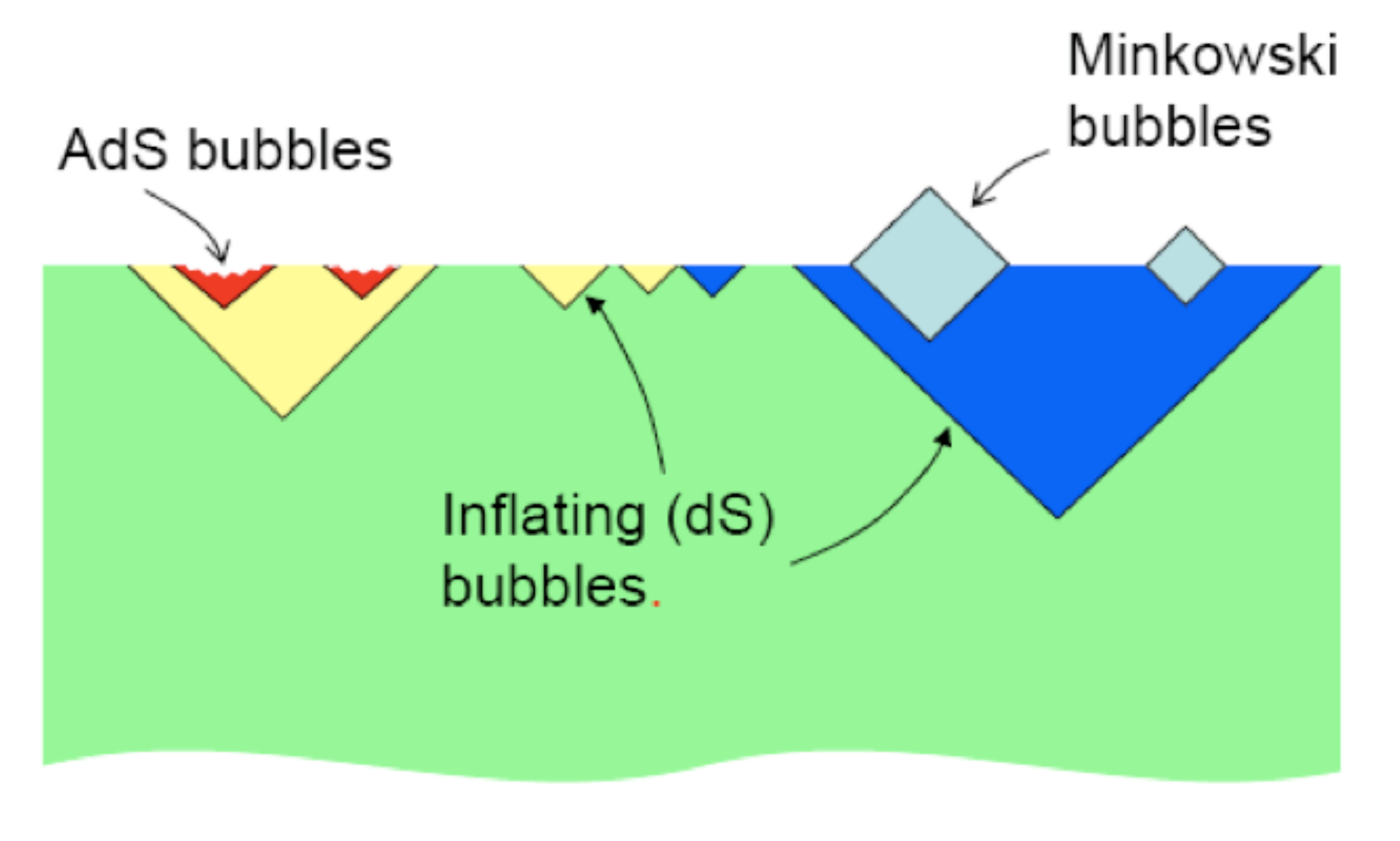}
\caption{Causal diagram of the inflationary multiverse.  The vertical direction represents time, and the horizontal direction is space.  Bubbles of different types nucleate and expand close to the speed of light.  Bubbles with positive vacuum energy (dS bubbles) inflate eternally.} 
\end{center}
\end{figure}

It has occasionally been suggested that spacetime does not really terminate at crunches in AdS bubbles.  Spacetime singularities may eventually be resolved in the fundamental theory of Nature, so that AdS crunches will become nonsingular.  The standard description of AdS regions will still be applicable at the
initial stages of the collapse, but when the density and/or curvature get sufficiently high, the dynamics would change, resulting in a bounce.  Scenarios of this sort have been discussed in the 1980's in the context of the so-called maximum curvature hypothesis \cite{Markov0,Markov1,Markov2}, and more recently in the context of pre-big-bang scenario \cite{Veneziano}, ekpyrotic and cyclic models \cite{Khoury,cyclic}, higher derivative models \cite{Biswas:2005qr},  loop quantum cosmology \cite{Ashtekar}, models involving ghost fields \cite{Peter,Allen,Xue}, ghost condensation \cite{Creminelli,Lin}, galileon models \cite{Easson,Cai}, and holographic ideas \cite{Brustein}.  

Because of the extreme (probably near-Planckian) energy densities reached near the bounce, the crunch regions are likely to be excited above the energy barriers between different vacua, so transitions to other vacua are likely to occur \cite{Piao1}.  If large inhomogeneities develop inside the bubble, then different parts of the crunch region may transit to different vacua, separated by domain walls \cite{watcher}.  The resulting global structure of spacetime is illustrated in Fig.2.  

Apart from their effect on the global structure, an additional motivation for studying AdS bounces comes from the measure problem.  This is the problem of assigning probabilities to different kinds of events in the multiverse.  Each kind of event occurs an infinite number of times in the course of eternal inflation, and  probabilities sensitively depend on how these infinities are regulated.  In a recent paper \cite{watcher} it has been suggested that AdS bounces may help to resolve the problem: the modified global structure of spacetime allows for a well defined probability measure.

\begin{figure}
  \includegraphics[width=16cm]{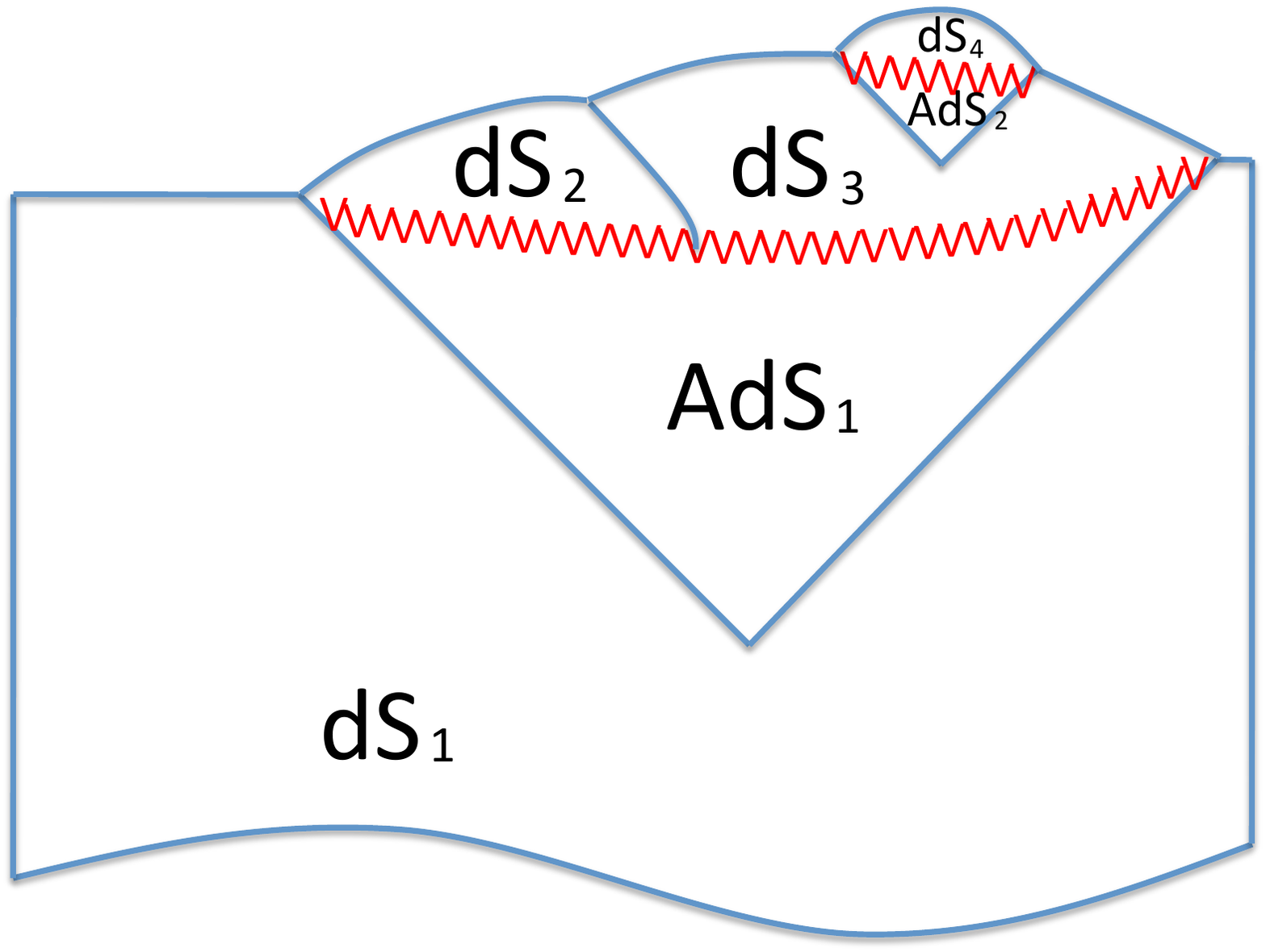}
  \caption{Causal diagram of a multiverse in which AdS crunches are non-singular and are followed by bounces.}
\end{figure}

The dynamics of AdS bounces and of transitions to other vacua has been discussed in a simple model by Piao \cite{Piao1,Piao2}. Our goal in the present paper is to provide a more detailed analysis.  In particular, we shall account for the effects of spatial curvature, which can be very significant in vacuum bubbles.  We shall also discuss different mechanisms of amplification of quantum fluctuations inside the bubble and whether or not they can lead to fragmentation of the bubble universe into regions of different vacua after the bounce.

For the present analysis we shall adopt a modification of Friedmann evolution equation with a correction proportional to the square of the density, such that the Hubble rate vanishes at some critical density $\rho_{max}$. The equation we shall use is similar to the effective Friedmann equation derived in Loop Quantum Cosmology (LQC) \cite{Ashtekar}.
The LQC effective equations have been extensively studied and have been shown to yield non-singular bouncing solutions.  Most of this work has focused on spatially flat ($K=0$) FRW models, although extensions to $K=\pm 1$ models have also been recently obtained \cite{Ashtekar2,Ashtekar4,Ashtekar3,Vandersloot,Szulc}. Here, we will be interested in the $K=-1$ case.  For reasons to be explained in Sec.~IV, we find the LQC effective equation for this case somewhat problematic, and we will use a slightly different form of it. We emphasize that this modified equation was not derived from LQC, and we use it only as a phenomenological model that yields non-singular bounces.

As we shall see, the correction terms to the Friedmann equation are negligible almost all the way to the bounce and almost immediately after the bounce.  Our main goal in this paper is to study the physical processes during the rapid contraction and re-expansion.  In particular, we shall be interested in the behavior of Higgs-like scalar fields that can be responsible for transitions between different vacua. Away from the bounce, the corrections to General Relativity should be small, and we shall use the standard theory of cosmological perturbations. 
Given that our current understanding of the dynamics {responsible for} the bounce is rather poor, we shall not attempt to address the behavior of perturbations during the time when the classical Einstein's equations do not hold, and simply assume that they are not significantly {modified} during this brief period.

The paper is organized as follows.  {In the next Section we review the Coleman-DeLuccia formalism which will be used to set up the initial conditions inside the bubble.  Bubble evolution starting with these initial conditions is discussed in Section III.  The modified Friedmann equation is introduced in Section IV and is then applied to study a non-singular bounce, both analytically and numerically.  Quantum fluctuations of the scalar field and some of their amplification mechanisms are discussed in Section V.  Finally, our conclusions are summarized and discussed in Section VI.}

\section{Coleman-de Luccia bubbles}

We shall consider a simple vacuum landscape with a single scalar field $\phi$ and a potential $V(\phi)$ having some dS and some AdS minima.  The potential is of a generic form 
\beq
V(\phi)=\lambda \eta^4 F(\phi/\eta),
\label{Vphi}
\eeq
with a dimensionless self-coupling $\lambda\ll 1$ and a characteristic energy scale $\eta\ll 1$. 
Starting with $\phi=\phi_F \sim \eta$ in an inflating dS vacuum, we consider quantum tunneling through a barrier to an adjacent AdS vacuum at $\phi=0$.   A prototypical double-well potential is illustrated in Fig.~3.  We shall now review the description of the bubble in the standard Coleman-DeLuccia formalism \cite{CdL}. 

\begin{figure}[t]
\begin{center}
\includegraphics[width=12cm]{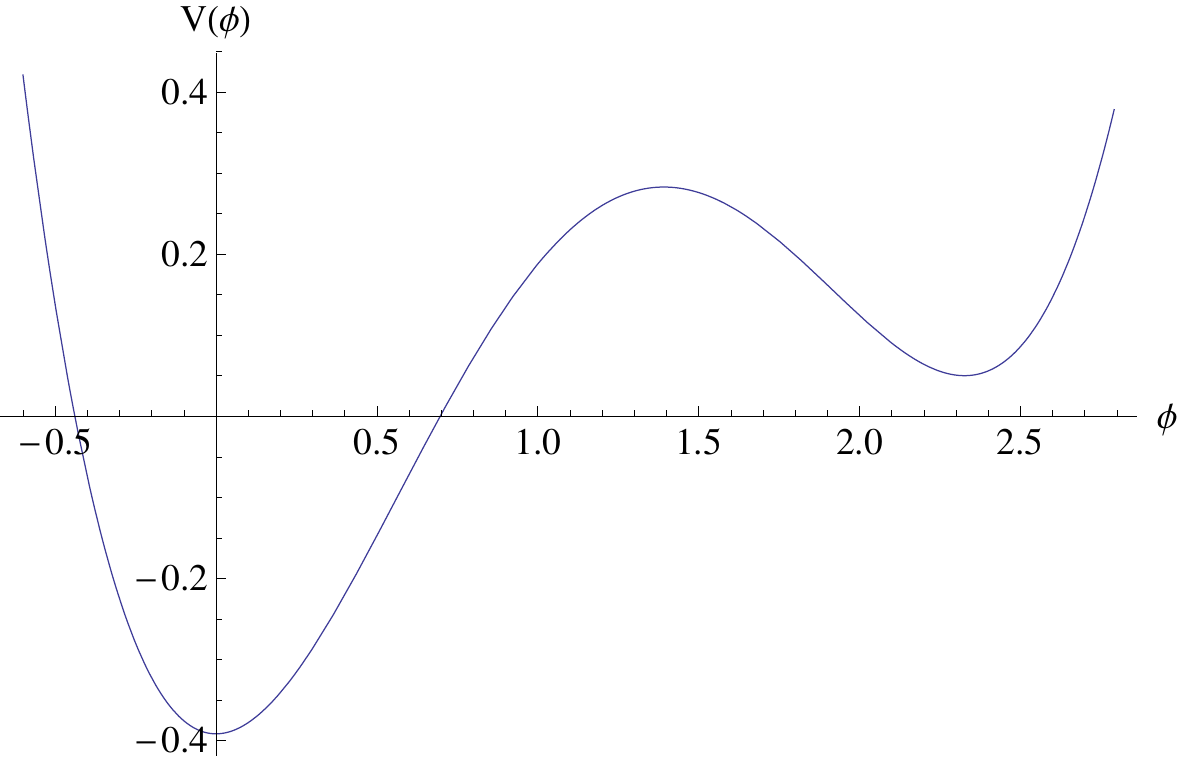}
\caption{A prototypical double-well potential.} 
\end{center}
\end{figure}

Disregarding perturbations, the interior of the bubble is described by a hyperbolic FRW metric,
\beq
ds^2 = -dt^2 + a^2(t)(d\xi^2 + \sinh^2 \xi d\Omega^2)
\eeq
and a homogeneous scalar field $\phi(t)$.  The functions $\phi(t)$ and $a(t)$ satisfy the equations
\beq
{\ddot\phi}+3\frac{\dot a}{a}{\dot\phi} +V'(\phi) = 0 , 
\label{phieq}
\eeq
\beq
{\dot a}^2 + K = \frac{1}{3}\rho a^2,
\label{aeq}
\eeq
where $K=-1$ is the curvature parameter, the energy density $\rho$ is given by
\beq
\rho = \frac{1}{2}{\dot\phi}^2 + V(\phi),
\label{rho}
\eeq
and we use Planck units where $8\pi G=1$.

The initial conditions for $a$ and $\phi$ can be obtained from the instanton -- a compact, $O(4)$-symmetric solution of the Euclidean Einstein and scalar field equations, which can be found from (\ref{phieq}),(\ref{aeq}) by analytic continuation $\zeta=it$, $\psi=i\xi$.  The instanton metric is given by
\beq
ds^2 = d\zeta^2 + b^2(\zeta) (d\psi^2 + \sin^2 \psi d\Omega^2) ,
\eeq
where $b(\zeta)=ia(-i\zeta)$ and the scalar field is $\phi=\phi(\zeta)$.  The coordinate $\zeta$ varies in the range $0<\zeta<\zeta_m$, and the functions $b(\zeta)$ and $\phi(\zeta)$ satisfy the regularity conditions
\beq
b(0) = b(\zeta_m) = 0, ~~~ b'(0)=-b'(\zeta_m) = 1 ,
\label{abc}
\eeq
\beq
\phi'(0) = \phi'(\zeta_m) = 0.
\label{phibc}
\eeq
They translate into the following initial conditions for $a$ and $\phi$ at $t=0$:
\beq
a(0)=0, ~~~~ {\dot a}(0)=1 ,
\label{abc}
\eeq
\beq
\phi(0)=\phi_0, ~~~~ {\dot\phi}(0)=0 ,
\label{phibc}
\eeq
where $\phi_0 = \phi(\zeta = 0)$.  These initial conditions ensure that the "big bang" at $t=0$ is a non-singular null surface.

The instanton solution $\phi(\zeta)$ also provides the field profile of the bubble at nucleation, with $\phi(0) = \phi_0$ being the scalar field value at the bubble center.  An important special case is that of a thin bubble, when the bubble radius is large compared to the thickness of the bubble wall, and $\phi_0$ is close to the AdS vacuum.  This is obtained when the energy difference between the two vacua is small compared to the height 
of the barrier between them \cite{Coleman},\footnote{We assume that the bubble is small compared to the de Sitter horizon and that gravitational effects on the tunneling are unimportant.}
\beq
q\equiv \frac{V_F -\rho_v}{V_m-V_F}\ll 1 ,
\label{thinwall}
\eeq
Here, $\rho_v \equiv V(0) <0$ is the AdS vacuum energy density, $V_F\equiv V(\phi_F)$, $V_m \equiv V(\phi_m)$, and
$\phi_m$ corresponds to the peak of the barrier.  In fact, our numerical calculations show that $\phi$ tends to tunnel almost all the way to the AdS minimum even when $q\sim 1$, and in order to have $\phi_0\sim\eta$, one has to get deep into the thick wall regime, $q\gg 1$.  To illustrate this, we plotted in Fig.~4 the quantity
\beq
r\equiv \frac{V_0-\rho_v}{V_F-\rho_v} ,
\label{r}
\eeq
where $V_0 \equiv V(\phi_0)$, as a function of $q$ for a quartic potential.  The effects of gravity have been ignored in Fig. 4.\footnote{Note that with a suitable rescaling of the scalar field and of spacetime coordinates, the quartic potential model without gravity includes only one independent parameter \cite{Sarid}.} 

\begin{figure}[t]
\begin{center}
\includegraphics[width=12cm]{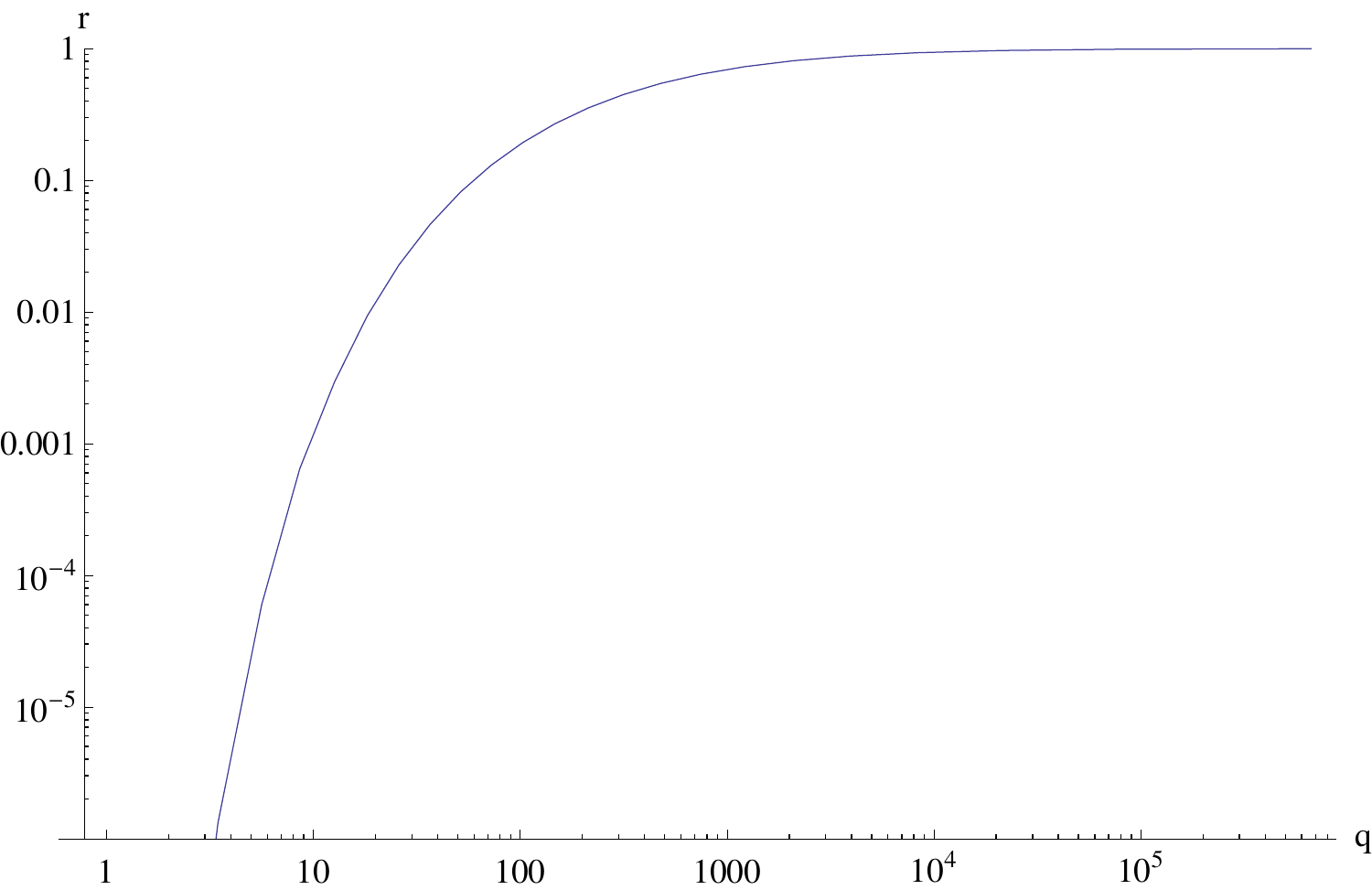}
\caption{The relationship between r and q for a quartic potential { in flat space. The value of $r$ is very small, say $r\lesssim .01$, unless we consider values of $q$ which are deep into the thick wall regime, $q\gtrsim 100$}.} 
\end{center}
\end{figure}


\section{Bubble evolution}

A detailed discussion of cosmological evolution with negative potentials has been given by Felder, Frolov, Kofman and Linde in \cite{FFKL}.  This paper provides important insights; however, it is focused mainly on flat models with $K=0$, and its conclusions do not directly apply to our case.  We shall consider two limiting regimes of evolution, corresponding to $r\ll 1$ and $r\sim 1$, which we shall refer to as thin wall and thick wall regimes, respectively.  { Unless stated othewise, we shall assume throughout the paper that there is no inflation inside the bubble; so $|V'/V(\phi_0)| \gtrsim 1$.}


In the thin wall regime, we have $r\ll 1$, and the initial value of $\phi$ is close to the AdS minimum, $|\phi_0|\ll\eta$.  However, we do not assume that $q\ll 1$ (see the discussion at the end of Sec.~II).  Instead, we shall assume that the model (\ref{Vphi}) has no small parameters apart from $\lambda$ and $\eta$.  Then the form of the potential near the minimum should be
\beq
V(\phi)\approx \rho_v +\frac{1}{2}m^2\phi^2 ,
\label{quadratic}
\eeq
with
\beq
\rho_v \sim -\lambda\eta^4
\eeq
and
\beq
m^2 \sim \lambda\eta^2 .
\eeq
  
The bubble universe is initially curvature dominated,
\beq
a(t)\approx t ,
\label{Milne}
\eeq
and the field $\phi$ is overdamped, $\phi\approx {\rm const}$.  At $t\sim m^{-1}$, $\phi$ begins to oscillate with a frequency $\omega \sim m$.  The total energy density can then be represented as 
\beq
\rho = \rho_v + \rho_m ,
\eeq
where the ``matter" component $\rho_m = \frac{1}{2}({\dot\phi}^2+m^2\phi^2)$ accounts for the field oscillations.  
At the onset of oscillations,
\beq
\rho_{m0}\sim r(V_F -\rho_v) \sim r|\rho_v| \ll |\rho_v|,
\label{rhom0}
\eeq
where $r$ is from Eq.~(\ref{r}).  The effective matter equation of state is that of a pressureless dust, so the matter density decreases as $\rho_m(t)\propto a^{-3}(t)$ in the course of expansion.  

As long as the density of matter is negligible compared to $\rho_v$, the evolution is approximately that of an empty AdS universe, 
\beq
a(t)\approx H_v^{-1} \sin (H_v t) ,
\label{empty}
\eeq
where
\beq
H_v=\left(|\rho_v| /3\right)^{1/2}.
\eeq
The expansion ends at $t\approx \pi /2H_v$ and is followed by contraction.  In the contracting phase, the matter energy density grows and becomes, on a similar timescale, $\sim \rho_{m0}$.  At this point, $a\sim m^{-1}$, $H=|{\dot a}/a| \sim m$, and the oscillating phase ends.  At later times, the evolution of $\phi$ is driven by the anti-damping effect of contraction, and the mass term $m^2\phi$ in Eq.~(\ref{phieq}) can be neglected:
\beq
{\ddot\phi}+3\frac{\dot a}{a}{\dot\phi} = 0 . 
\label{phieq1}
\eeq
This has the solution 
\beq
{\dot\phi} \propto a^{-3}.
\label{dotphi}
\eeq  
The matter density is then
\beq
\rho_m = \frac{1}{2}{\dot\phi}^2 = C a^{-6} ,
\label{rhoa6}
\eeq
which corresponds to the equation of state $P=\rho$.  The constant $C$ can be estimated from the condition $\rho_m(a\sim m^{-1})\sim \rho_{m0}$,
\beq
C\sim m^{-6}\rho_{m0} \sim r(\lambda\eta)^{-2}.
\label{C}
\eeq

Close to the big crunch, $(t_c -t)\ll t_c$, where $t_c =\pi/ H_v$, the scale factor (\ref{empty}) is
\beq
a(t)\approx (t_c - t) ,
\label{curvdom}
\eeq
and Eqs.~(\ref{dotphi}), (\ref{rhoa6}) give
\beq
\phi(t) \approx \sqrt{\frac{C}{2}} (t_c - t)^{-2}
\label{phicurvdom}
\eeq
Note that even though the field grows very fast, the mass term $m^2\phi$ in the field equation remains small compared to the other terms in (\ref{phieq1}), which grow like $(t_c-t)^{-4}$.  However, nonlinear terms in (\ref{phieq}) may eventually become important.  For example, a quartic term in the potential would give a cubic term in the field equation, which would grow like $(t_c-t)^{-6}$.  When this term becomes dominant, the growth of the field (\ref{phicurvdom}) terminates, the field gets reflected from the potential wall and starts moving in the opposite direction.  

On the other hand, if the potential is bounded from above, $|V(\phi)|<V_{max}$, it may remain unimportant all the way to the crunch.  Once the kinetic energy gets much larger than $V_{max}$, the potential plays negligible role in the dynamics.  In this paper we shall consider only such bounded landscapes.  Specifically, we shall assume that $|F(x)|\lesssim 1$ in Eq.~(\ref{Vphi}), so $V_{max}\sim\lambda\eta^4 \ll 1$.

At $a\sim C^{1/4}$, the kinetic energy density (\ref{rhoa6}) comes to dominate and the curvature term in (\ref{aeq}) becomes negligible:
\beq
\frac{{\dot a}^2}{a^2} = \frac{1}{3}\rho.
\label{aeq1}
\eeq
The solution of Eq.~(\ref{aeq1}) with $\rho$ from (\ref{rhoa6}) is
\beq
a(t) = (3C)^{1/6} (t_c - t)^{1/3} ,
\label{asol}
\eeq
where the value of $t_c$ may be slightly different from that in (\ref{curvdom}), (\ref{phicurvdom}). 
Substituting this back to (\ref{aeq1}) and using $\rho={\dot\phi}^2/2$, we find
\beq
\phi = \phi_* + \sqrt{\frac{2}{3}} \ln \frac{t_c - t_*}{t_c-t} ,
\label{phisol}
\eeq
where $\phi_*$ is the value of $\phi$ at the beginning of the kinetic energy dominated (KED) period, 
$t_c - t_* \sim C^{1/4}$.  From Eq.~(\ref{phicurvdom}), we find $\phi_* \sim 1$.   Since the field $\phi$ grows only logarithmically in the KED regime, nonlinear terms in the field equation remain negligible if they did not become important before.


In the thick wall regime, we have $q\gg 1$, $r\sim 1$, and we expect $\phi_0\sim\eta$ and $\rho_{m0}\sim |\rho_v|$.  In this range of $\phi$, Eq.~(\ref{quadratic}) may not be a good approximation, and the initial evolution will depend on the shape of the potential.  As before, the expansion is initially curvature dominated (CD).  Field oscillations that begin after this CD period are not generally harmonic, and the effective equation of state may significantly differ from $P=0$.  If the potential is sufficiently fine-tuned, there may even be a period of slow-roll inflation prior to oscillations.  But eventually the oscillations do begin, and their amplitude is damped by the expansion, which brings us to the regime with $\rho_m\ll |\rho_v|$ and $\rho_m\propto a^{-3}$ discussed above.
 
In the contracting phase, the oscillation amplitude will grow, and when it gets $\sim\eta$, we may again have a transition period where the field evolution is sensitive to the shape of the potential.  This will be followed by a CD period (during which the bubble universe will contract by a factor $\sim \eta^{-1}$) and then by a KED period.


The solutions (\ref{curvdom}), (\ref{phicurvdom}), (\ref{asol}), (\ref{phisol}) describe a universe collapsing to a singularity at $t=t_c$.  The collapse will be cut off if the singularity is resolved and replaced by a bounce, as we discuss in the next Section.

\section{Non-singular bounce}

\subsection{Modified Friedmann equation}

In this Section, we introduce a heuristic model for the dynamics of the bounce.  This consists of a simple modification of the Friedmann equation (\ref{aeq}),
\beq
\frac{{\dot a}^2}{a^2} = \left(\frac{\rho}{3} - \frac{K}{a^2}\right)
\left(1-\frac{\rho}{\rho_c} \right). 
\label{aeq2}
\eeq
such that when $\rho$ reaches the value $\rho_c$, we have ${\dot a}=0$, and the universe bounces.
Here. $\rho_c$ is a free parameter and we assume that $\rho_c\lesssim 1$, so that the semiclassical picture of spacetime does not break down.

Eq.~(\ref{aeq2}) is very similar to the effective equation
\beq
\frac{{\dot a}^2}{a^2} = \left(\frac{\rho}{3} - \frac{K}{a^2}\right)
\left(1-\frac{\rho}{\rho_c} +\frac{3K}{a^2 \rho_c} \right) ,
\label{aeq3}
\eeq
which has been {derived} for $K=0,-1$ in Loop Quantum Cosmology (LQC) \cite{Ashtekar2,Ashtekar4,Vandersloot,Szulc}.  (The
equation for $K=+1$ is more complicated \cite{Ashtekar3}, but here we are only interested in $K=-1$.)    
In (\ref{aeq3}), $\rho_c$ is the characteristic density at which LQC corrections become important.  In the LQC formalism, the value of $\rho_c$ is related to the so-called Barbero-Immirzi parameter \cite{Ashtekar}.  
As we mentioned in the introduction, the form of the equation (\ref{aeq3}) is somewhat problematic.  The initial conditions inside the bubble require $a\to 0$ at a finite $\rho = V_0$.  However, it is easy to see that this is inconsistent with Eq.~(\ref{aeq3}) for $K=-1$.  There is also another problem.  For $K=0$ and $\rho=0$, Eq.~(\ref{aeq3}) has Minkowski space, $a={\rm const}$, as a solution, as it should.  On the other hand, for $K=-1$ and $\rho=0$,
the Milne universe, $a=t$, is not a solution.  Instead, Eq.~(\ref{aeq3}) has a solution
\beq
a^2 = a_0^2 +t^2,
\eeq
where $a_0^2 = 3/\rho_c$.  This, however, is not satisfactory, since the Milne universe can be obtained from Minkowski space by a coordinate transformation.  
It should be noted that the derivation of the LQC effective equation (\ref{aeq3}) for $K=-1$ in Refs.~\cite{Vandersloot,Szulc} is not at the same level of rigor as that for $K=0,+1$ in \cite{Ashtekar2,Ashtekar3}.\footnote{We thank Parampreet Singh for a discussion of this issue.} It is conceivable, therefore, that a more rigorous treatment would yield a different equation.  
Here, we shall simply adopt the model given by Eq.~(\ref{aeq2}), which does not have the above mentioned problems.

\subsection{Curvature dominated bounce}
 
We shall distinguish the cases where the bounce occurs in the CD and in the KED regimes.  On the approach to a CD bounce, the scale factor is given by (\ref{curvdom}), and from (\ref{rhoa6}) the energy density is
\beq
\rho_m \approx \frac{C}{(t_c-t)^6} .
\eeq
The bounce occurs when $\rho_m \approx \rho_c$, that is, at $(t_c-t)\approx (C/\rho_c)^{1/6}$.  Requiring that the transition to the KED regime has not occurred prior to the bounce, $(t_c-t) \gtrsim C^{1/4}$, we obtain the condition 
\beq
C\lesssim \rho_c^{-2} ,
\label{Ccond}
\eeq
or, using the estimate (\ref{C}) for $C$,
\beq
r \lesssim \left(\frac{\lambda\eta}{\rho_c}\right)^2 . 
\eeq

From Eq.~(\ref{phicurvdom}), the scalar field displacement {in a CD bounce is}
\beq
\Delta\phi \approx {\sqrt{2}} (C\rho_c^2)^{1/6} \lesssim 1,
\label{Deltaphi}
\eeq
where we added a factor of 2 in the first step, to account for the field displacement before and after the bounce.

The duration of a CD bounce can be estimated as  
\beq
\Delta t \sim (C/\rho_c)^{1/6} .
\eeq
In order for our semiclassical treatment to apply, we have to require that $\Delta t\gtrsim 1$, or $C\gtrsim \rho_c$.

\subsection{Kinetic energy dominated bounce}

For $C\gg \rho_c^{-2}$, the bounce is in the KED regime.  In this case, we can find an analytic solution for $a(t)$ near the bounce.  Neglecting the curvature term in Eq.~(\ref{aeq2}) we have
\beq
\frac{{\dot a}^2}{a^2} = \frac{\rho}{3} \left(1-\frac{\rho}{\rho_c}\right) ,
\label{aeq4}
\eeq
and with $\rho$ from Eq.~(\ref{rhoa6}) the solution is \cite{Piao1}
\beq
a(t) = \left(\frac{C}{\rho_c} + 3C(t_c - t)^2\right)^{1/6} .
\label{bouncesol}
\eeq

The bounce occurs at 
\beq
a_{min} = (C/\rho_c)^{1/6} \gg 1 
\eeq
and has the duration $\Delta t \sim \rho_c^{-1/2}$.  The field
displacement during the KED period can be estimated by setting 
$|t_c - t|\sim \rho_c^{-1/2}$ in Eq.~(\ref{phisol}) and multiplying by 2 to account for the periods before and after the bounce.  This gives 
\beq
\Delta\phi \sim \frac{1}{\sqrt{6}} \ln (C\rho_c^2) \gtrsim 1.
\eeq

\subsection{Multiple bounces}

For both CD and KED bounces, the kinetic energy density of the field $\phi$ at the bounce is very large, $\rho_m \approx \rho_c$.  As the universe expands after the bounce, the field slows down and eventually settles into one of the minima of the potential.  If it is a dS vacuum, the field oscillations about the minimum are rapidly damped, and the bubble universe enters an epoch of inflationary expansion.  This epoch will continue forever, with bubbles of other vacua occasionally being formed amidst the inflating background.

If the field settles into the basin of attraction of an AdS vacuum, then the bubble universe goes through an oscillating period followed by a bounce, as outlined above, but with one important caveat.  The initial conditions for $\phi$ are now set not by the instanton, but by the details of the flight from the preceding AdS vacuum.  
Since $\phi$ flies to the new vacuum over the barrier, the initial oscillation amplitude is typically large, corresponding to $r\sim 1$.

\subsection{A toy landscape}

To illustrate this discussion, we now present a numerical
  solution of the evolution equations (\ref{phieq}), (\ref{aeq2}) for
  a toy landscape with a potential $V(\phi)$ shown in Fig.~5.
This includes several dS and AdS vacua and has $\eta\sim 0.1$ and
$\lambda\sim 0.6$.  We use $\rho_c = 0.01$.

\begin{figure}[t]
\begin{center}
\includegraphics[width=12cm]{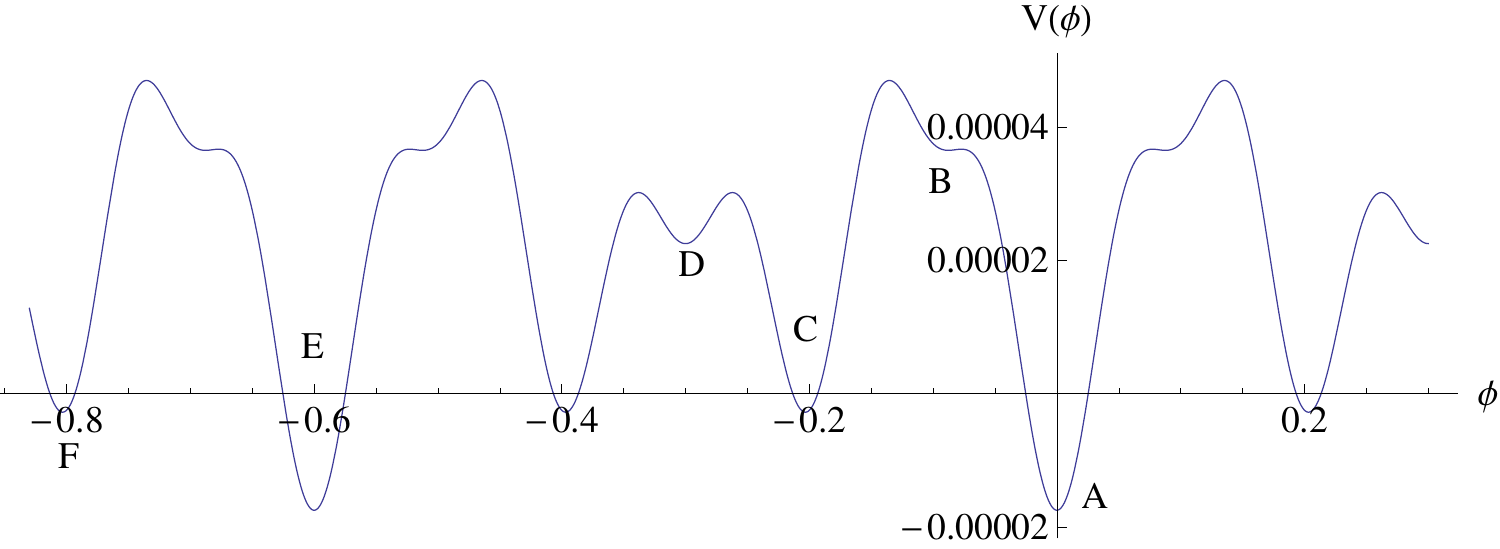}
\caption{A toy landscape with several vacua labeled by $A, B, C, D$, among which vacua $A$ and $C$ are AdS vacua, and vacua $B$ and $D$ are dS vacua.} 
\end{center}
\end{figure}

We assume that the field $\phi$
starts in the dS vacuum labeled $B$ and tunnels to the AdS vacuum $A$.  
To find the initial 
value $\phi_0$ after the tunneling, we solve the Euclidean field
equations for the instanton; this gives $\phi_0 =-0.027$.  
The barrier is rather 
flat, so we are in the thick wall regime, $q = 311.8$, and the initial
energy density $V_0$ is well above the vacuum level, $r = 0.36$.    

The subsequent field evolution is plotted in Fig.~6.  At first, the
field oscillates about the AdS minimum $A$, with the amplitude of
oscillation decreasing during the expansion and increasing during the
contraction period.  Then, at the time of the bounce, the field shoots
very rapidly, flying over the initial vacuum $B$, then over another
AdS vacuum $C$, and lands in the dS vacuum labeled $D$.

\begin{figure}[t]
\begin{center}
\includegraphics[width=12cm]{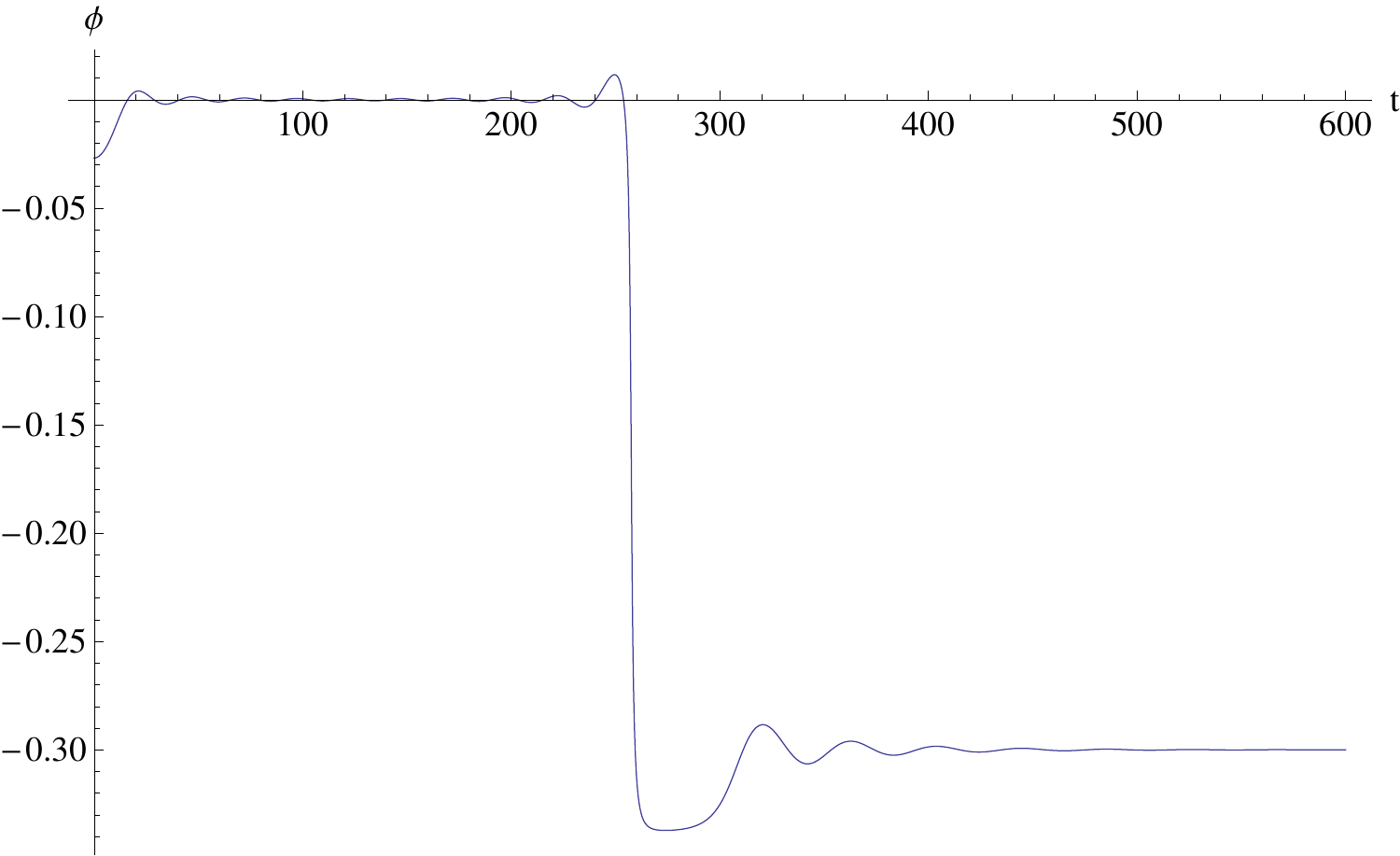}
\caption{Evolution of the field. {The field starts in vacuum $B$, from which it tunnels to $A$. After some oscillations, the field picks up speed on its way to the bounce and shoots back, flying over the original dS vacuum $B$,
the adjacent AdS vacuum $C$, and ends up landing in the inflating vacuum $D$}. } 
\end{center}
\end{figure}

\begin{figure}[t]
\begin{center}
\includegraphics[width=12cm]{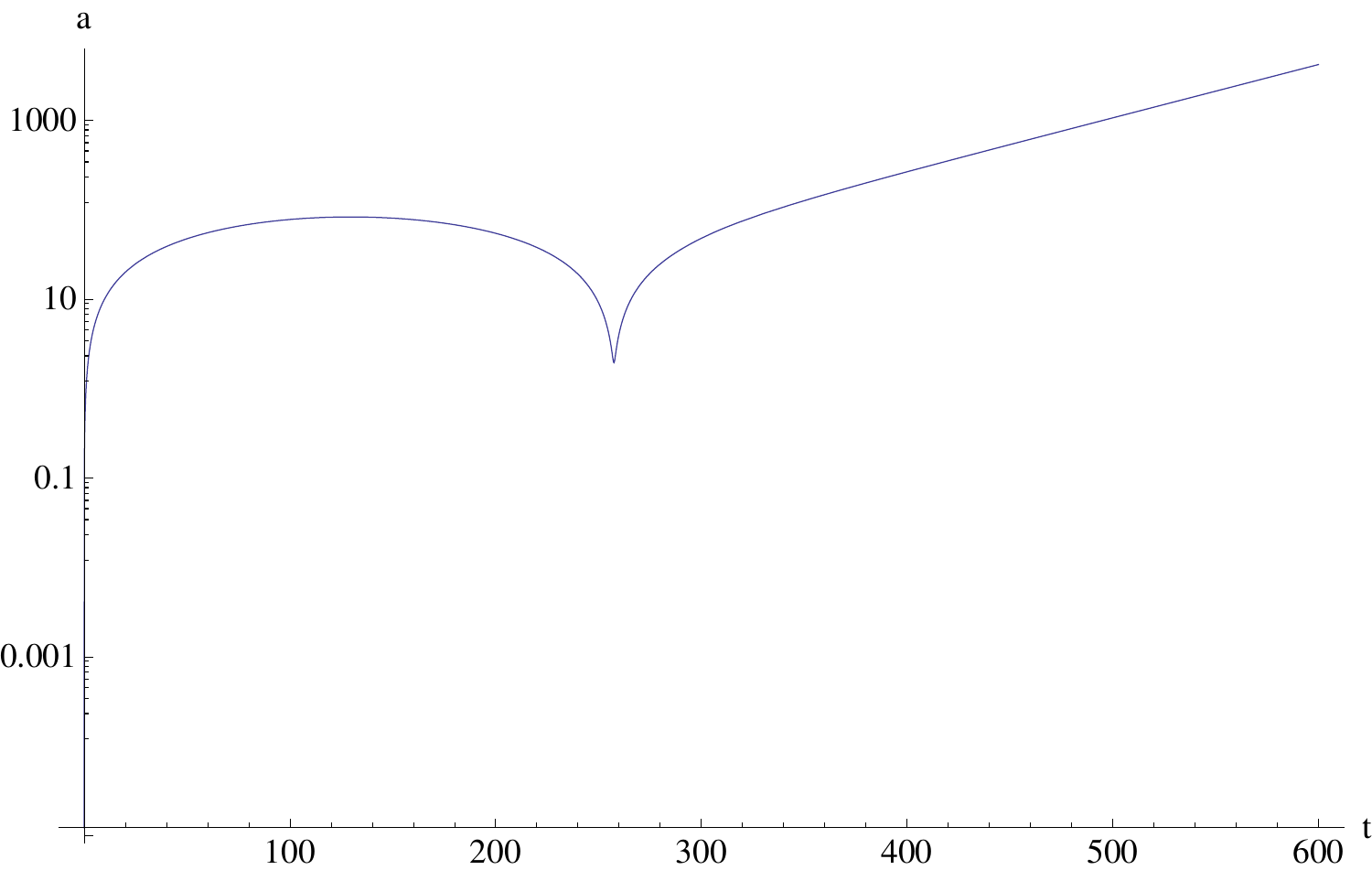}
\caption{Evolution of the scale factor. {Expansion and recollapse in the AdS vacuum A are followed by a bounce.  After the field $\phi$ lands in vacuum $D$, inflation resumes.}} 
\end{center}
\end{figure}

The scale factor evolution is shown in Fig.~7.  The straight line
after the bounce represents exponential expansion in the dS vacuum
$D$.

Fig.~8 shows the relative magnitude of the density and curvature
terms, $|\rho|/3$ and $1/a^2$, in Eq.~(\ref{aeq2}).  We see that in this
example curvature dominates throughout the bounce.  The sharp downward
peaks in the density term correspond to the moments when the density changes
sign.  (Note that we have plotted the absolute value of the density in
Fig.~8.)  After the field settles into the inflationary vacuum $D$,
the curvature term rapidly declines.

\begin{figure}[t]
\begin{center}
\includegraphics[width=12cm]{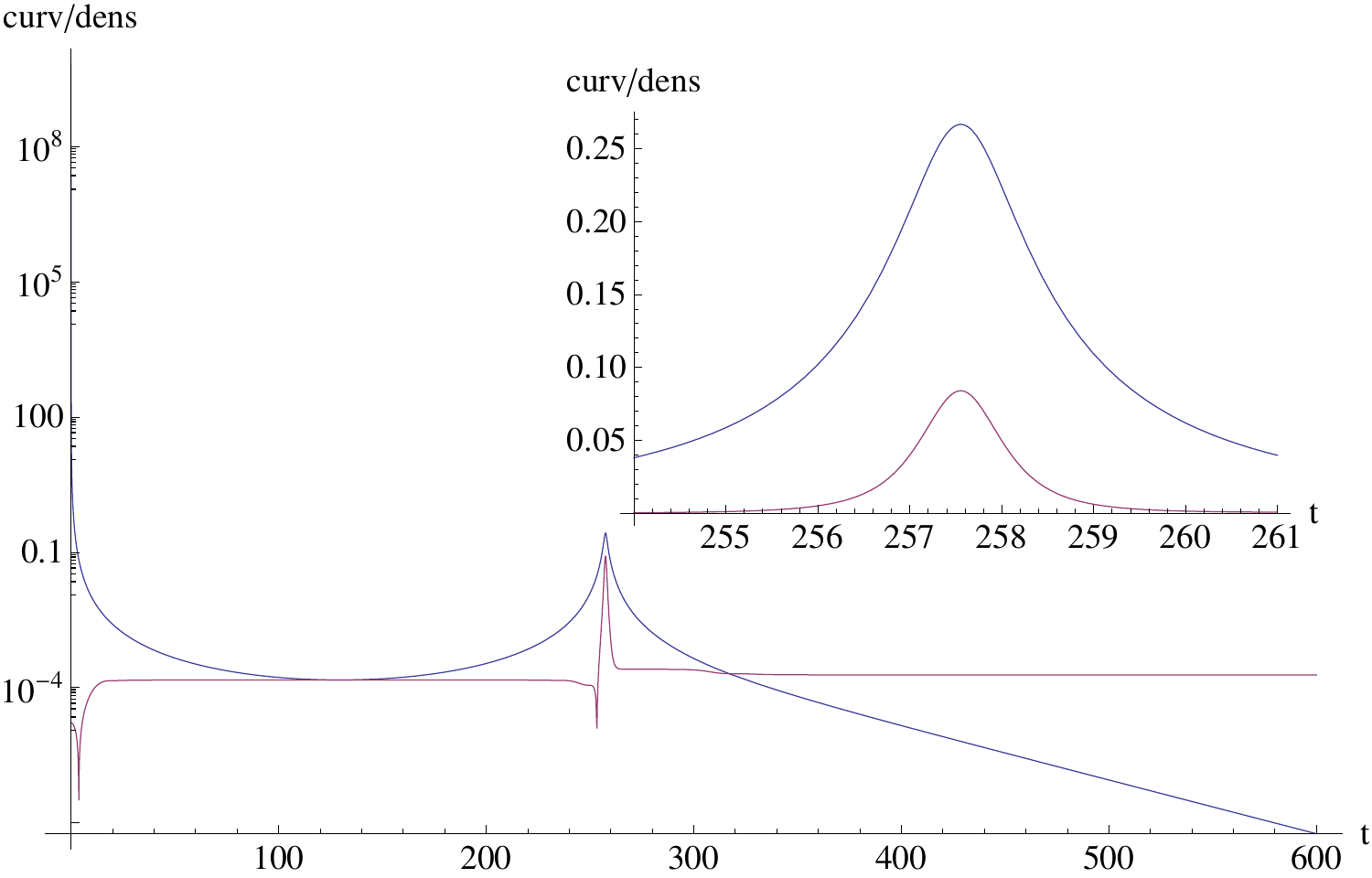}
\caption{The relative magnitude of terms in Eq.~(\ref{aeq2}). The blue line is the curvature term and the red line is the {absolute value of the density}. The insert plot shows the magnitude of these two terms near the bounce.} 
\end{center}
\end{figure}

\begin{figure}[t]
\begin{center}
\includegraphics[width=12cm]{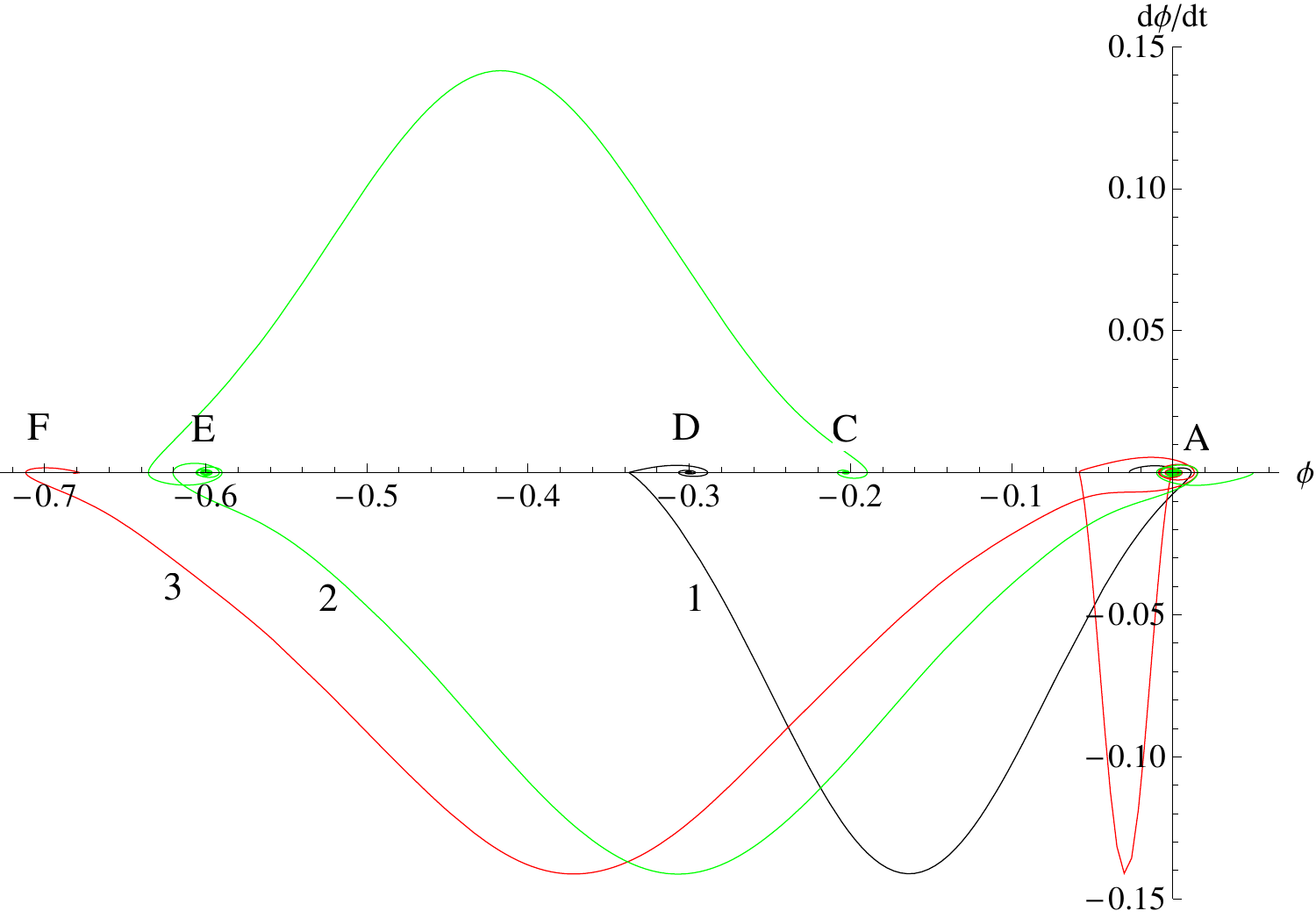}
\caption{Phase space trajectories with different initial conditions in the landscape of Fig.~5.} 
\end{center}
\end{figure}

The value of the parameter $C$ can be found by fitting our numerical
solutions to Eq.~(\ref{rhoa6}); this gives $C = 0.53$.  Note that the
condition (\ref{Ccond}) is satisfied, so we should be in the curvature
dominated regime, in agreement with Fig.~8.  The field displacement
estimate (\ref{Deltaphi}) gives $\Delta\phi = 0.27$, which is close to
the numerical displacement in Fig.~6.

{To illustrate different kinds of field evolution, we have plotted in Fig.~9 some phase space trajectories in the $\{\phi,{\dot\phi} \}$ plane, starting with different initial conditions in the attraction basin of vacuum $A$.  The initial conditions at $t=0$ resulting from tunneling from the vacuum $B$ to vacuum $A$ are $\phi_0=-0.0269$, ${\dot\phi}_0 =0$; the corresponding phase space trajectory (labeled by $1$ in the figure) is shown by the black line in the figure.  After oscillating about vacuum $A$ at the origin, the field  goes through the bounce and settles in the dS vacuum $D$.   Note that the characteristic width of the potential well of vacuum $A$ is $\Delta\phi \sim 0.05$, so the starting point of this trajectory is relatively far from the potential minimum.  

As we discussed in Section IV.D, the initial conditions can be more or less arbitrary when the vacuum $A$ is reached after a bounce in some other AdS vacuum.  The phase trajectory labeled $2$, shown by the green curve in the figure, starts from $\phi_0 = 0.05$ with ${\dot\phi}_0 =0$.  After the bounce, the field lands in the basin of another AdS vacuum $E$, and following yet another bounce, it transits to a dS vacuum $C$.  The red curve labeled $3$ starts very close to the potential minimum, at $\phi_0 = -0.0001$.  (This could be the initial condition after a thin-wall tunneling.)   Here, the field $\phi$ returns to vacuum $A$ after the first bounce and then shoots to a dS vacuum $F$.
}

\section{Fluctuations}

Quantum fluctuations in the scalar field $\phi$ and in the metric will necessarily cause deviations from homogeneity and isotropy.  These fluctuations can be amplified by various mechanisms.  If they get sufficiently large, they may cause different parts of the bubble to transit to different vacua after the bounce.  Amplification of quantum fluctuations has been extensively discussed in the context of inflationary scenario and of ekpyrotic and cyclic models.  In our context, the same basic mechanisms will operate, but their effectiveness and some qualitative features will be different.

In most of the earlier studies of cosmological quantum fluctuations,
the goal was to find the fluctuation spectrum and use it to derive
predictions for structure formation and for CMB anisotropies.  Our
goal here will be simpler.  We assume that the AdS bubble formed somewhere in the chain of
transitions that led to our present vacuum, so the fluctuations
generated in the bubble are not directly related to present
observations.  { We would like to clarify the question of} whether the fluctuations grow
large enough to cause the bubble interior to split into regions of
different vacua after the bounce.  This will typically happen when the
scalar field fluctuations become $O(\eta)$ on the length scale $\gtrsim m^{-1}$.

The field $\phi$ can be represented as
\beq
\phi({\bf x},t) = \phi(t) + \delta\phi({\bf x},t),
\eeq
where $\phi(t)$ is the homogeneous component and the fluctuation $\delta\phi$ can be expanded into modes \cite{Garriga}
\beq
\delta\phi({\bf x},t) = \sum_{klm} \left[ \phi_k(t) Y_{klm}({\bf x}) a_{klm} + c.c.\right] .
\label{deltaphi}
\eeq
Here, ${\bf x}$ stands for coordinates on a unit 3-hyperboloid and the functions $Y_{klm}$ and $\phi_k$ satisfy the equations
\beq
\left(\nabla^2 + k^2 \right) Y_{klm} ({\bf x}) = 0 ,
\eeq
\beq
{\ddot\phi}_k+3\frac{\dot a}{a} {\dot\phi}_k +\frac{k^2}{a^2}\phi_k  +V''(\phi)\phi_k=0,
\label{phik}
\eeq
where $\nabla^2$ is the Laplacian on the 3-hyperboloid {and we have assumed that the effect of metric perturbations on the evolution of modes $\phi_k$ can be neglected.}.

The mode spectrum includes a continuous part, $1<k<\infty$, as well as some discrete modes with $k^2 < 1$, {which are known as supercurvature modes}.  In particular, { in the approximation where self-gravity of the bubble can be neglected,} there is always a mode with $k^2=-3$ { at the bottom of the spectrum, which comes from the fluctuations in the shape of the bubble wall  \cite{GV92}} and corresponds to a position-dependent time shift of the homogeneous field.  One can eliminate this time shift by a coordinate transformation; then the $k^2=-3$ mode reappears as a tensor perturbation with wavelength comparable to the hyperbolic curvature radius \cite{Garriga}. { When gravity of the bubble is included, the same effect can be seen directly in the spectrum of tensor modes \cite{YST,GMST}. The wavelength of the $k=1$ mode of the continuous spectrum is also comparable to the curvature radius, and all modes with $k>1$ have shorter wavelengths.}

{Supercurvature modes with $k^2 \ll 1$ have wavelengths much larger than the curvature scale. In single field models, the existence of such modes would require a finely tuned potential. On the other hand, they are frequently encountered in multifield models \cite{quasiopen}, when light fields are present. In this case, such modes  provide an additional contribution to the matter energy density inside the bubble, which varies from place to place with a large coherence length of order $k^{-1}$ in units of the curvature scale. For the sake of simplicity, in this paper we concentrate on single field models, so we shall only consider the bubble fluctuation supercurvature mode.}

We shall now discuss various amplification mechanisms.

\subsection{Tachyonic amplification}

This mechanism operates when the homogeneous field $\phi(t)$ is in the range where $V''(\phi)<0$.  
The effective mass of the perturbations in Eq.~(\ref{phik}) is then tachyonic, $m_{eff}^2 = V''(\phi)<0$, and the modes with physical momenta 
\beq
p = k/a < |m_{eff}| 
\label{p}
\eeq
get exponentially amplified \cite{Felder,Hamazaki}.  This could happen at the early stages after bubble nucleation if the initial value $\phi_0$ is sufficiently close to the top of the barrier, so that $V''(\phi_0) < 0$.  Clearly, this situation is possible only for a thick wall bubble with $q\gg 1$, since otherwise $V''(\phi_0)\approx m^2 >0$.

The corresponding amplification factor is roughly given by 
\beq
F\sim \exp(\mu \Delta t), 
\label{factor}
\eeq
where $\mu \equiv |V''(\phi_0)|$ and $\Delta t$ is the time it takes for the field $\phi$ to cross the tachyonic region.
For a generic potential of the form (\ref{Vphi}), in the absence of fine-tuning, we expect $\mu\sim m\sim \sqrt{\lambda}\eta$.  The Hubble parameter is $H\sim \sqrt{\lambda}\eta^2 \ll m$, so the Hubble damping is negligible, and $\Delta t\sim m^{-1}$.  Hence, this mechanism is not likely to amplify fluctuations by more than an order of magnitude.

This seems to be at odds with the discussion in Felder et al \cite{Felder}, who argued that tachyonic preheating can be extremely efficient and can damp all the energy of the homogeneous field into short-wavelength perturbations within a single oscillation. The reason for this apparent discrepancy is that Felder et al considered fine-tuned, inflationary potentials and assumed that the field starts close to the maximum of the potential, where the Hubble damping is strong.

A stronger tachyonic instability can develop after the bounce, as the field $\phi$ travels to another vacuum.
Initially the velocity ${\dot\phi}$ is very large, and the field can
fly over a number of vacua without being much affected by the
potential.  However, as $\phi$ approaches its final destination, it
slows down.  It slows even more as it climbs over the potential
barrier on its way to the final vacuum.  While it is near the top of
the barrier, $V'' <0$, and tachyonic amplification can be significant.

This effect is illustrated in Figs.~10 and 11, where we have plotted,
respectively, $m_{eff}^2(t)$ and the amplification factor $F$ for the
$k=1$ mode in the landscape of Fig.~5.  $m_{eff}^2(t)$
undergoes rapid oscillations as the field flies over the vacua $B$ and
$C$, but at this time the field velocity is very large, and the
effective mass term has little effect on its dynamics.  A prolonged
period of negative $m_{eff}^2$ occurs when the field approaches the
final vacuum $D$.  Fluctuations grow by a factor $\sim 50$ during this
period. 

\begin{figure}[t]
\begin{center}
\includegraphics[width=12cm]{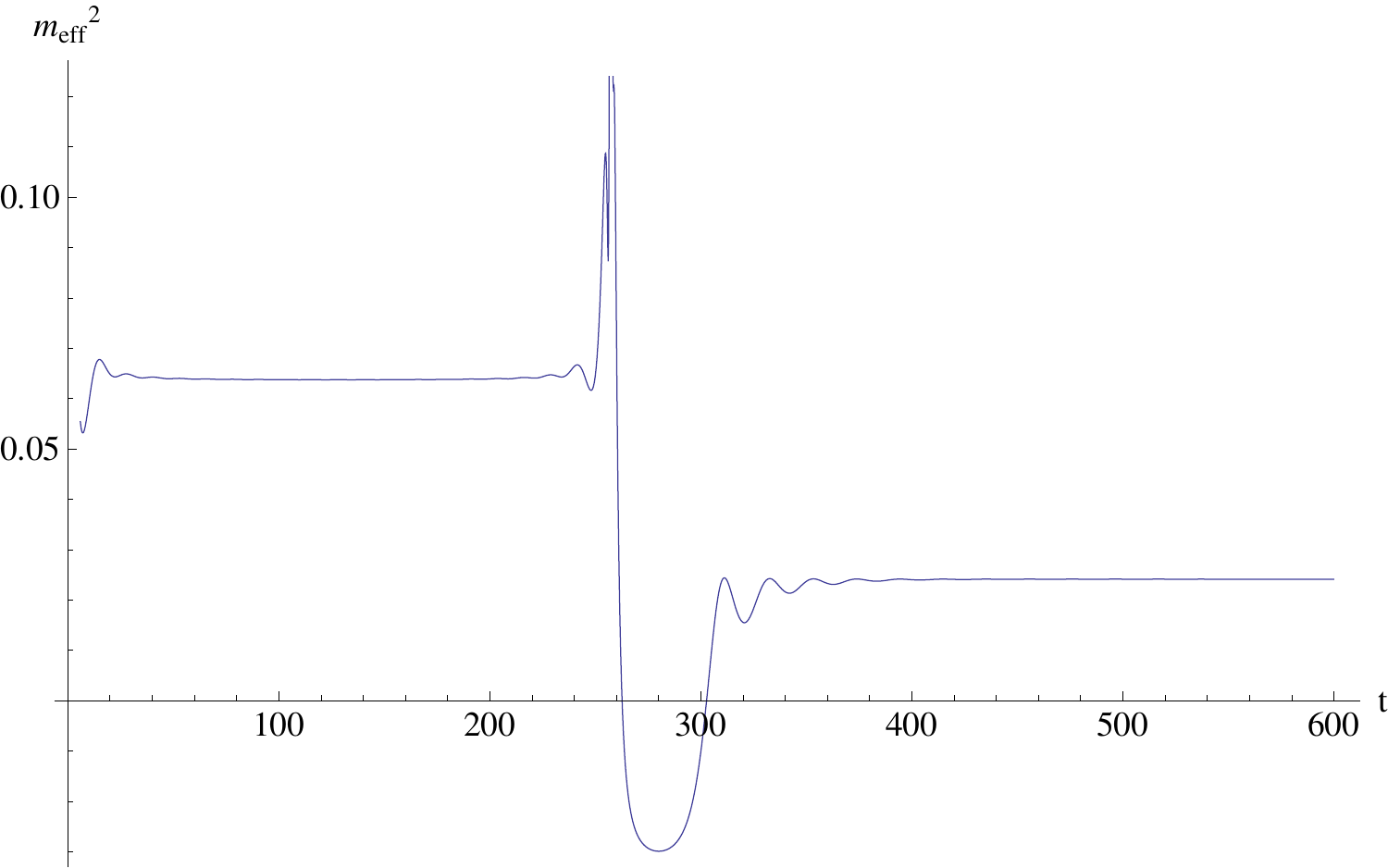}
\caption{The effective mass squared as a function of time.} 
\end{center}
\end{figure}

\begin{figure}[t]
\begin{center}
\includegraphics[width=12cm]{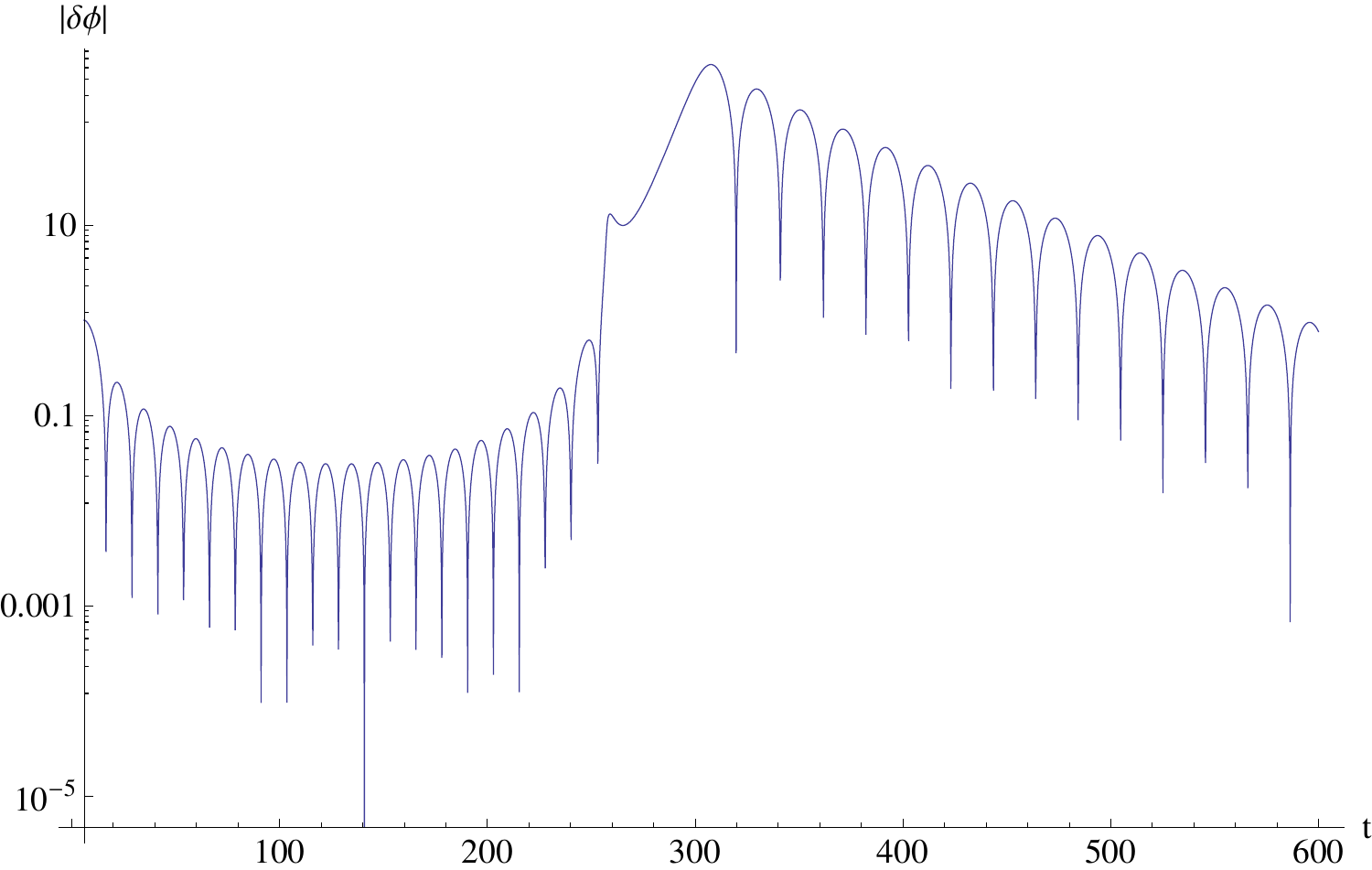}
\caption{Amplification of $k=1$ mode of the fluctuations.} 
\end{center}
\end{figure}

Once the field settles into vacuum $D$, the fluctuations are quickly
damped by the de Sitter expansion.  But if instead the field ended up
in an AdS vacuum, the fluctuations could be further amplified,
as discussed in Sec.~IV.E.

The initial amplitude of perturbations on scale $l\gtrsim 2\pi/\mu$ can be estimated as \cite{Felder}
\beq
\langle \delta\phi^2 \rangle \sim \int_0^\mu \frac{kdk}{4\pi^2} =\frac{\mu^2}{8\pi^2} ,
\eeq
which gives the characteristic amplitude
\beq
\delta\phi_{in} \sim \frac{\mu}{2\pi} .
\label{phiin}
\eeq
Assuming that the time $\Delta t$ of crossing the tachyonic region is smaller than the Hubble time, the wavelength of perturbations remains nearly constant during the amplification process.  The enhanced amplitude after the amplification is
\beq
{\delta\phi_f} \sim \frac{\mu F}{2\pi} \sim \frac{\sqrt{\lambda}}{2\pi}\eta F.
\label{phif}
\eeq
This shows that if the coupling $\lambda$ is not too small, the
fluctuations can become comparable to the field vev $\eta$ even for
modest values of the amplification factor $F$. 


\subsection{Parametric resonance}

Apart from the tachyonic instability, the modes $\phi_k$ can also be amplified by parametric resonance during the oscillating phase \cite{Kofman}.  Parametric resonance can also damp the energy of the homogeneous mode of $\phi$ into short-wavelength fluctuations of other fields. 

The effectiveness of this mechanism depends on the parameter $Q\sim \lambda \Phi^2 /m^2$, where $\Phi$ is the amplitude of oscillations of $\phi$.  For $Q\gg 1$ the resonance is broad and amplification is very efficient, while for $Q\ll 1$ the resonance is narrow and amplification is inefficient.  
A broad resonance is typically achieved when the oscillation phase follows slow-roll inflation and the initial value of $\phi$ is close to the maximum of a very flat potential.  In our case, $\Phi\lesssim\eta$ and $Q\lesssim 1$; 
thus we expect the effect to be only marginally efficient for a generic $V(\phi)$.  In the numerical examples that we studied, we saw no evidence for a significant resonant amplification.

\subsection{Domain wall fluctuations, tensor modes and the BKL regime}

As we already mentioned, domain wall fluctuations yield a position-dependent time shift of the homogeneous field.  A time shift by itself does not change the evolution history and the final destination of the field $\phi$.  However, the position dependence results in anisotropies, which can have a significant impact on the dynamics.  It is well known that anisotropies tend to grow dramatically near a big crunch singularity: the shear terms in Einstein equations grow as $a^{-6}$, which is faster than the usual matter and radiation densities.  When anisotropy becomes large, the universe enters a Belinsky-Khalatnikov-Lifshitz (BKL) regime, with a stochastic succession of Kasner epochs \cite{BKL}.   In our case, however, the matter density in the kinetic energy dominated (KED) regime is also $\propto a^{-6}$, so the evolution may be different.  If anisotropies do not become large prior to KED, then the contraction remains approximately isotropic and the BKL regime is never reached \cite{BKL,Steinhardt}.

The evolution of anisotropies can be studied using the tensor mode representation of wall fluctuations.  At early times after bubble nucleation, the metric perturbation is given by \cite{Garriga}  
\beq
h \sim \frac{{\dot a}}{a} (R_0\sigma)^{-1/2} \sim \eta^{-1} \frac{{\dot a}}{a} ,
\label{h}
\eeq
where $R_0\sim m^{-1}$ is the initial radius of the bubble and $\sigma\sim \lambda^{1/2} \eta^3$ is the bubble wall tension.  The subsequent evolution of $h$ is described by the usual equation for tensor perturbations 
\cite{Mukhanov} with $K=-1$ and $k^2=-3$,
\beq
{\ddot h}+3\frac{\dot a}{a} {\dot h} +\frac{1}{a^2} h = 0.
\label{heq}
\eeq

During the period from bubble nucleation till the onset of the KED phase, the scale factor is approximately given by (\ref{empty}), and {Eq.~(\ref{heq}) has two linearly independent solutions
\beq
h_1(t) = \cot X, 
\label{cot}
\eeq
\beq
h_2(t) = \cot X \ln [\tan(X/2)] + (\sin X)^{-1} ,
\label{cotlog}
\eeq
where $X = H_v t$.   The initial condition (\ref{h}) selects $h \propto h_1(t)$.  (Note that the asymptotic behavior of the two modes near the crunch is $h_1(t)\propto (t_c -t)^{-1}$ and $h_2(t)\propto (t_c-t)^{-1} \ln(t_c-t)$, so the dominant mode at $t\to t_c$ is $h_2(t)$.  The difference, however, is only in the logarithmic factor.)}    

According to the discussion in Sec.~III, the KED epoch is present only if $C \sim r(\lambda\eta)^{-2}\gg 1$.  It  begins at $t_*$ with $t_c-t_* \sim C^{1/4}$.  At that time,
\beq
h_*\equiv h(t_*) \sim \eta^{-1}(t_c-t_*)^{-1} \sim C^{-1/4} \eta^{-1}.
\label{h*}
\eeq
This can be large or small, depending on the magnitudes of $C$ and $\eta$. For $h_*\ll 1$, the contraction is approximately isotropic all the way to the bounce.  For $h_* > 1$, the stochastic BKL regime sets in prior to $t_*$.   Different curvature-radius-size regions will then follow completely different evolution, so the metric will become highly inhomogeneous.\footnote{In the CD bounce scenario, the bounce occurs at $(t_c-t_B)\approx (C/\rho_c)^{1/6}$, when $h_B \sim \eta^{-1} (\rho_c/C)^{1/6}$.  If $h_B >1$, this indicates that the BKL regime is reached before the bounce.}

The dynamics in the high-energy regime near the bounce is ultralocal: gradient terms in Einstein and scalar field equations are small compared to time derivatives, so the evolution locally follows that of a homogeneous (but anisotropic) universe.  The field equation for $\phi$ in the KED regime is then
\beq
\frac{d}{dt} (\sqrt{-g} {\dot\phi})=0.
\label{phieq3}
\eeq
This shows that the evolution of $\phi$ depends only on the determinant of the metric $g$.  In the BKL regime, the behavior of the metric is highly irregular, but $g(t)$ is a regular function
\beq
g(t)\propto (t_c-t)^2 ,
\eeq
independent of position \cite{BKL}.  The density of matter in the KDE regime is $\rho \approx {\dot\phi}^2/2 \propto g^{-1}$, so if the universe is nearly homogeneous at the beginning of the BKL epoch, the density and the field $\phi$ will remain homogeneous.  This seems to suggest that the bounce will occur everywhere at nearly the same time and that the solution of Eq.~(\ref{phieq3}) for $\phi(t)$ will be the same as in the homogeneous case, given by Eq.~(\ref{phisol}).
We note, however, that the evolution of the metric determinant $g$ during brief transition periods between Kasner regimes will differ in different locations.  Also, the ultralocal approximation will break down at the beginning and at the end of the BKL epoch.  Both of these effects can in principle lead to significant scalar field and density fluctuations, resulting in transitions to different vacua and in black hole formation.  The dynamics of the bounce may also be influenced by the anisotropy \cite{Ashtekar,Singh}. Analysis of these issues is beyond the scope of the present paper, but intuitively one expects that the scalar field will be rather inhomogeneous after a BKL epoch.

\section{Summary and discussion}

We have explored some of the cosmological consequences of the assumption that Einstein's equations get modified by quantum effects, resulting in resolution of spacetime singularities.  Our focus has been on the big crunch singularities in AdS bubbles in the multiverse scenario, and we have adopted a simple modification of Friedmann equation (\ref{aeq2}), combined with a generic scalar field model.  With this modification, the singular big crunch is replaced by a non-singular bounce, which occurs when the density of matter reaches certain critical value $\rho_c\lesssim 1$.  A comoving region in the multiverse will then go through a succession of dS and AdS epochs.  Our goal in this paper was to study the dynamics of a typical AdS bounce. 

Some of the qualitative features of AdS bounces have been outlined by Piao in Ref.~\cite{Piao1}.  The main limitation of his analysis was that he assumed spatial curvature to be negligible, while we show here that the evolution of a typical AdS bubble is curvature dominated almost all the way to the bounce.  We find nevertheless that Piao's quailtative conclusion still holds: the scalar field gets excited above the barriers separating different vacua, so AdS bubbles are likely to transit to other vacua after the bounce.  

The character of the bounce dynamics is determined by the parameter $C$ defined in Eqs.~(\ref{rhoa6}) and (\ref{C}).  For $C\lesssim \rho_c^{-2}$, the dynamics is curvature dominated (CD) all the way through the bounce, while for $C\gg \rho_c^{-2}$ curvature domination is followed by a kinetic energy dominated (KED) regime.  The field displacement is $\Delta\phi\lesssim 1$ and $\Delta\phi\gtrsim 1$ for CD and KED bounces respectively.  If the field $\phi$ has a geometric origin, as in string theory landscape, {then its full range is 
often restricted not to exceed the Planck scale by order of magnitude.  (For a discussion of this restriction and of the possible ways to circumvent it, see Ref.~\cite{Silverstein}.)}  This implies that AdS bounces can cause transitions to distant parts of the landscape. 

We have also discussed amplification mechanisms of quantum fluctuations in AdS bubbles.  Scalar field fluctuations can be amplified by tachyonic instability and by parametric resonance.  We find that these mechanisms are much less efficient than in models of slow-roll inflation.  {If the fluctuations remain small, then the entire bubble transits to the same new vacuum (apart perhaps from some rare localized regions).
But in some models, especially when the scalar field is relatively strongly coupled, the amplification can be strong enough to cause significant fractions of the bubble volume to transit to different vacua.}  

If the new vacuum after the bounce is AdS, the whole story is repeated, with another bounce and transition to another vacuum (or set of vacua).  {Scalar field fluctuations developed during the first bounce can get further amplified in subsequent bounces, so the probability for the bubble to split into multiple regions of different vacua is enhanced.}\footnote{Johnson and Lehners \cite{JohnsonLehners} discussed models of eternal inflation, which allow for the possibility of cyclic bubble universes with non-singular bounces.  For a homogeneous bubble, this would lead to an infinite repetition of the same evolution history inside the bubble.  However, perturbations may accumulate from one cycle to the next and may eventually disrupt the cyclic behavior.  Refs.~\cite{Piao2} and \cite{Zhang} argue that this indeed happens for the curvature perturbations.}

Bubble wall fluctuations induce tensor modes of metric perturbations inside the bubbles.  Depending on the parameters of the model, this can result in strong anisotropy in a contracting AdS bubble and can lead to a BKL regime of alternating Kasner epochs.  Then the universe emerges after the bounce in a highly inhomogeneous state.  Density inhomogeneities can lead to black hole formation, and scalar field inhomogeneities can induce fragmentation of the bubble universe into regions of different vacua.

{Our analysis has focused on a `typical' bounce that occurs in a `generic' potential, having no small parameters other than the coupling $\lambda$ and the characteristic energy scale $\eta$ (which are not necessarily assumed to be very small).  However, in a string theory landscape with $\sim 10^{500}$ vacua, 
there will also be AdS minima with `fine-tuned' potentials, where these generic conditions are not satisfied.}  For example, there will be AdS bubbles where the effective mass of $\phi$ is initially very small.  Then there may be a period of slow-roll inflation inside the bubble.   This will reduce the importance of spatial curvature and can make it completely negligible.  The subsequent evolution of the bubble may then be similar to that of our own universe, except in the end the dark energy density is negative.  If the matter dominated epoch is long enough for structure to develop, it will lead to large local variations in the matter density, {resulting in a} different field evolution during and after the bounce.   

{While this paper was being written, we learned of a closely related work by B.~Gupt and P.~Singh \cite{GuptSingh}, which is expected to appear on the arXiv simultaneously.}

\subsection*{Acknowledgements}

J.G. is grateful to the Tufts Institute of Cosmology for warm hospitality during the preparation of this work. This work was supported in part by grant AGAUR 2009-SGR-168, MEC FPA 2010-20807-C02-02, CPAN CSD2007-00042 Consolider-Ingenio 2010 (J.G.) and by the National Science Foundation (grant PHY-1213888) and the Templeton Foundation (A.V.).  We are grateful to Abhay Ashtekar, Sugumi Kanno, Matt Kleban and Parampreet Singh for very useful and stimulating discussions.

\end{document}